\documentclass[conference]{IEEEtran}
\IEEEoverridecommandlockouts
\usepackage{cite}
\usepackage{amsmath,amssymb,amsfonts}
\usepackage{algorithmic}
\usepackage{graphicx}
\usepackage{textcomp}
\usepackage{xcolor}
\usepackage{enumitem}
\usepackage{tabularx} 
\usepackage{longtable}
\usepackage{xspace}
\usepackage[utf8]{inputenc}
\usepackage{booktabs}
\usepackage{tabularx}
\usepackage{float}

\usepackage{url}
\def\BibTeX{{\rm B\kern-.05em{\sc i\kern-.025em b}\kern-.08em
    T\kern-.1667em\lower.7ex\hbox{E}\kern-.125emX}}
\begin{document}

\newcommand{\ie}{\textit{i.e.,}\xspace}
\newcommand{\eg}{\textit{e.g.,}\xspace}
\newcommand{\etc}{\textit{etc.}\xspace}
\newcommand{\etal}{\textit{et al.}\xspace}

\newcommand\sanny[1]{\textcolor{red}{#1}}
\newcommand\gdf[1]{\textcolor{blue}{#1}}
\newcommand\gs[1]{\textcolor{orange}{#1}}

\title{PrediHealth: Telemedicine and Predictive Algorithms for the Care and Prevention of Patients with Chronic Heart Failure.
}
\author{\IEEEauthorblockN{Giuseppe De Filippo$^*$, Simranjit Singh$^*$,  Gianpiero Sisto$^*$, Marco Mazzotta$^\#$,Gianvito Mitrano$^\#$, \\
Claudio Pascarelli$^\#$, Gianluca Fimiani$^+$, Simone Romano$^+$, Mariangela Lazoi$^\#$, Marina Garofano$^+$, \\ Alessia Bramanti$^+$} 
\IEEEauthorblockA{\textit{$^*$MediNet S.r.L}, \textit{$^\#$University of Salento}, \textit{$^+$University of Salerno}}
}

\maketitle

\begin{abstract}
The management of chronic heart failure presents significant challenges in modern healthcare, requiring continuous monitoring, early detection of exacerbations, and personalized treatment strategies. 
This paper presents the preliminary results of the PrediHealth research project conducted in this context. Specifically, it aims to address the challenges above by integrating telemedicine, mobile health solutions, and predictive analytics into a unified digital healthcare platform. We leveraged a web-based IoT platform, a telemonitoring kit with medical devices and environmental sensors, and AI-driven predictive models to support clinical decision-making. 
The project follows a structured methodology comprising research on emerging CPS/IoT technologies, system prototyping, predictive model development, and empirical validation. 
\end{abstract}


\section{Introduction}
CardioVascular Diseases (CVDs) remain the leading cause of mortality, morbidity, hospitalization, and disability in many countries, imposing a significant burden on healthcare systems. For example, direct healthcare costs related to CVDs account for approximately 65\% of all public health expenditures in Italy, with pharmaceutical expenses constituting an additional 25\%. Heart Failure (HF) is particularly concerning as it is one of the fastest-growing cardiovascular conditions, primarily due to the increasing life expectancy of the population. The Diagnosis-Related Group (DRG) 127, which pertains to HF, ranked as the second most frequent DRG in Italy, with 223,240 hospitalizations reported in 2015. Furthermore, within 30 days of hospital discharge, the mortality rate for HF patients stands at 11\%, with readmission and rehospitalization rates of 14\% and 3\%, respectively~\cite{MinistryHealthReport}. Things are similar in different countries, where HF represents a significant healthcare challenge. In Germany, for example, HF has become the most common reason for hospitalization. Notably, in 2017, HF was the leading cause of in-hospital death, accounting for 8.9\% of all cases, with a higher rate in East Germany (65 deaths per 100,000 population) compared to West Germany (43 deaths per 100,000 population)~\cite{neumann2009heart}. In the United Kingdom, HF is responsible for 1–2\% of total healthcare expenditure, reflecting its substantial economic impact~\cite{cowie2017heart}. 
In the United States, approximately 6 million individuals were living with HF between 2015 and 2018, with prevalence projected to increase by 46\% between 2012 and 2030~\cite{osenenko2022burden}.

Managing HF is challenging, and the situation often worsens when dealing with elderly and frail patients, who often suffer from multimorbidity and a reduced physiological reserve. These factors contribute to frequent hospitalizations and high mortality rates. These data underscore the global challenge posed by HF, emphasizing the need for effective management strategies to alleviate its burden on healthcare systems worldwide. Current healthcare models struggle to provide effective post-hospitalization management, leading to fragmented care, increased healthcare costs, and suboptimal patient outcomes. Addressing these issues requires an integrated, technology-driven approach that enhances patient monitoring, enables early detection of deteriorating conditions, and supports personalized interventions~\cite{rockwood1999brief}.

\begin{table}[t]
\centering
\caption{Project summary information.}\label{tab:project}
 \scalebox{0.75}{
\begin{tabular}{@{}rp{5.5cm}@{}} \toprule
\textbf{Name:} & PrediHealth \\ 
\textbf{Website:} & {\url{https://tinyurl.com/2mwbu7r4}} \\
\textbf{Duration:} & 15 months (from July 2024) \\
\textbf{Funding Agency:} & National Recovery and Resilience Plan, Mission 4, Component 2, ‘From Research to Business.’  \\
\textbf{Participants:} &  University of Salerno and University of Salento  \\
\textbf{Principal Investigators:} & Alessia Bramanti, Giuseppe Scanniello, and Mariangela Lazoi \\
\textbf{Funding Received} & 207,963.00 €\\
\textbf{Industrial Partner:} & MediNet S.r.L. (https://www.medinetsrl.eu) \\
\textbf{Technology Readiness Level (TRL)} & 6 \\
\textbf{Status:} & Ongoing \\ \bottomrule
\end{tabular}
}
\end{table}

In this paper, we present the PrediHealth research project (see Table~\ref{tab:project} for some additional information). This project focuses on using CPS (Cyber-Physical Systems) and IoT solutions\footnote{While CPS encompasses a broader range of networked systems, IoT specifically focuses on internet-enabled connectivity.} in smart healthcare and medical research. The project is conducted in collaboration with a software industrial partner operating in CPS and IoT fields and aims to facilitate the integrated hospital-territory-home management of HF patients. It leverages state-of-the-art technologies such as telemedicine, CPS/IoT devices (wearable and not), Artificial Intelligence (AI), and Machine Learning (ML). 

One of the primary goals of this paper is to share with the community the research underlying PrediHealth, the encountered challenges, and its designed and engineered solutions. Specifically, we provide here the following main contributions:
\begin{itemize}[leftmargin=2mm]
    \item A web-based IoT platform to securely use and exchange telemonitoring data and ensure integration with existing hospital electronic health records. To achieve this, we conducted a review of open-source IoT-based web applications, evaluating them against a set of criteria, including security, interoperability, scalability, and ease of integration. Based on this analysis, we selected the most suitable application and adapted it to meet the specific requirements of our project.
    
    \item A telemonitoring kit, comprising medical-grade wearable devices and environmental sensors, designed for real-time health tracking of HF patients. To ensure its effectiveness and suitability, we conducted a review of these devices/sensors, assessing them against a set of predefined criteria, including accuracy, reliability, interoperability, and patient comfort. 
    
    \item A decision-making support system, developed using AI/ML-powered algorithms to analyze patient-centered data, facilitating patients' stratification\footnote{It consists in categorizing individuals into different risk groups based on their health status, disease progression, or other factors.} and early intervention. The system is designed to identify HF risk patients who would benefit from continuous remote monitoring. The final goal is to determine which patients require a telemonitoring kit to be proactively enrolled in a personalized monitoring program.
    
\end{itemize}


\section{Background, Related Work, and Motivation} \label{sec:StateOfArt}
In this section, we review related work and background information relevant to the PrediHealth project and its foundational research, with a focus on (i)~Remote Patient Monitoring (RPM) and telemonitoring, (ii)~mobile health (m-health) solutions, and (iii)~predictive models for CVDs.

\textbf{RPM and Telemonitoring.} RPM and telemonitoring are closely related concepts. Specifically, RPM is a broad term that refers to the use of digital technologies to collect health data from patients outside traditional healthcare settings (\eg at home) and transmit it to healthcare providers. On the other hand, telemonitoring is a subset of RPM that specifically focuses on continuous or periodic remote monitoring of patients with chronic diseases, such as HF. Telemonitoring emphasizes real-time or near-real-time data collection, trend analysis, and early detection of worsening conditions to enable timely interventions. In short, RPM is a broader category that includes various remote data collection methods, while telemonitoring is more focused on continuous disease-specific monitoring and proactive healthcare interventions. 
Studies have demonstrated the effectiveness of RPM/telemonitoring in improving patient outcomes, particularly for chronic conditions like HF. In particular, Bashshur \etal~\cite{bashshur2014empirical} conducted a systematic review examining telemedicine's impact on chronic diseases, including HF, stroke, and chronic obstructive pulmonary disease. Their findings reveal reductions in mortality (by 15\% to 56\% in HF patients), hospital admissions, and healthcare costs. 
Similarly, Ma \etal~\cite{ma2022telemedicine} highlight the role of telemedicine consultation and telemonitoring in managing hypertension, diabetes, and rheumatoid arthritis. The study emphasizes the importance of long-term interventions.

\textbf{M-health.} m-health and telemonitoring are closely linked, as both leverage digital technologies to enhance remote patient care. M-health primarily focuses on mobile applications and wearable devices that enable real-time health tracking, patient engagement, and self-management. Telemonitoring extends this concept by integrating m-health data with broader healthcare systems.
M-health applications have shown considerable promise in enhancing patient engagement and promoting self-management, particularly for individuals with chronic diseases such as HF. Stampe \etal~\cite{stampe2021mobile} explored the use of m-health technologies in chronic disease management, emphasizing their role in improving patient-physician consultations. 
The researchers also observed that the use of these technologies helped in better decision-making by both patients and clinicians. 
Schmaderer~\etal~\cite{schmaderer2022mobile} conducted a pilot study focused on m-health self-management interventions for patients with HF. Their results indicated significant improvements in symptom management and overall health outcomes among patients who utilized m-health apps. Specifically, the study found a notable reduction in hospitalizations and readmissions, suggesting that m-health apps could be instrumental in managing HF symptoms and preventing exacerbations. 

\textbf{Predictive Models for CVDs.} Predictive AI/ML-powered models play a crucial role in managing chronic diseases like HF by enabling early detection, risk stratification, and personalized treatment for patients. These models analyze large datasets (\eg from wearable and environmental sensors) to support proactive interventions, reduce hospitalizations, and improve patient outcomes, making predictive modeling a key component of modern telemedicine and, more specifically, telemonitoring.
For example, Medhi \etal~\cite{medhi2024artificial} conducted a review by selecting 150 studies from databases such as PubMed, Google Scholar, and Cochrane Library. Their synthesis detailed how AI/ML-powered models---spanning classical methods like support vector machines, random forest, and decision trees to advanced deep learning approaches---can process large datasets from wearables, imaging, and other clinical sources to enable early detection, risk stratification, and personalized treatment of HF. Their work highlights the potential of these predictive models to reduce hospitalizations and improve patient outcomes and emphasizes the transformative role of AI in modern telemedicine and telemonitoring. Visco~\etal~\cite{visco2024explainable} developed a predictive model for cardiovascular diagnostics. Their study identified correlations between worsening HF and three key parameters: creatinine, systolic pulmonary artery pressure, and coronary artery disease. To enhance interpretability, they introduced a classifier based on genetic programming that achieved 97\% accuracy, outperforming traditional ML~models.

\textbf{Summary.} Existing literature supports the potential impact of integrating telemonitoring, m-health, and predictive models in managing chronic conditions like HF. The PrediHealth project builds on these insights by providing an integrated solution to improve the personalization of care, enhance patient engagement, and prevent exacerbations. Furthermore, the project contributes to the growing body of knowledge on the use of AI/ML models in healthcare, with a focus on HF.

\section{Objectives} \label{sec:objectives}

The high-level workflow of the PrediHealth project begins with the stratification of patients at risk of HF using an AI/ML-powered predictive model. Patients identified as at-risk are then equipped with wearable and environmental sensors, which are integrated into a telemonitoring system based on a web-based IoT platform. To achieve the overarching goal of PrediHealth, we have then defined and pursued the following operational objectives:

\textbf{O1: Develop an interoperable web-based IoT platform.} The objective is to develop an interoperable web-based IoT platform. The platform has to establish a secure, scalable, and standardized digital infrastructure for the collection, transmission, and analysis of health data from telemonitoring devices. This platform facilitates real-time monitoring of HF patients by integrating wearable and non-wearable medical devices as well as environmental sensors. Additionally, the platform has to incorporate advanced security mechanisms, including encryption and role-based access controls, to protect sensitive patient data and comply with relevant regulatory frameworks (\eg GDPR). The web-based architecture will also support scalability and flexibility, enabling future integration with different telemonitoring devices and services. 
A key aspect of O1 is to have a platform that supports HL7 FHIR (Fast Healthcare Interoperability Resources), a healthcare data exchange standard that enables interoperability across diverse (healthcare) ecosystems. By adhering to this standard, the platform developed in PrediHealth ensures compatibility with hospital information systems, electronic health records, and other telemedicine solutions.

\textbf{O2: Develop a telemonitoring kit for HF patients.} This objective focuses on the development of a comprehensive telemonitoring kit specifically designed for patients with HF, integrating advanced (wearable and not) medical devices and environmental sensors to enable continuous health monitoring. The kit ensures non-intrusive, user-friendly, and clinically effective monitoring. Key components of the kit include wearable medical devices, such as smartwatches and scales, capable of tracking critical physiological parameters like, for example, heart rate, weight, Heart Rate Variability (HRV), blood pressure, oxygen saturation (SpO2), and ElectroCardioGraphic (ECG) signals. Additionally, environmental sensors will monitor factors such as air quality, temperature, and humidity since these elements could affect cardiovascular health.
The devices of the telemonitoring kit will be interoperable with the platform developed in PrediHealth. 

\textbf{O3: Develop a multimarker score to support decision-making.} The goal of O3 is twofold: (i)~building AI/ML-powered models to facilitate the risk stratification of patients vulnerable to acute exacerbations or disease progression for HF (\ie identify patients at HF risk) and (ii)~defining thresholds and scores for risk factors associated with HF exacerbations of patients (those previously considered at HF risk) when using telemonitoring kits. As for the first point, a Decision Support System (DSS) developed using advanced AI/ML algorithms is designed to analyze patient-centered data and facilitate risk stratification. This system leverages historical medical records to generate an HF risk for each patient. That is, the primary function of this DSS is to identify HF risk patients who are most vulnerable to acute exacerbations or disease progression for HF. This allows making informed decisions about the patient--determining who would benefit the most from a telemonitoring program--and then allocating such a patient to that program. As for the second point, we aim to design and implement a multimarker score that integrates multiple parameters (both personal and environmental) to enhance clinical decision-making in HF management. This score takes into account real-time data from telemonitoring devices: IoT-enabled (wearable and not) medical devices and environmental sensors. The multimarker score is developed by identifying key biomarkers and risk factors associated with HF exacerbations, such as HRV, SpO2, respiratory rate, blood pressure fluctuations, body weight changes, and environmental conditions (e.g., air quality, humidity, and temperature). This aims to detect early signs of decompensation, enabling timely interventions and reducing the likelihood of emergency hospitalizations. In case the patient is not at home, only the data from  IoT-enabled medical devices will contribute to the score to support decision-making. 

\textbf{O4: Demonstrate and validate the proposed prototype.} We empirically evaluate the research solutions implemented in PrediHealth to ensure its technical feasibility, clinical relevance, and usability within real-world healthcare environments. Our validation will take place within THE (Tuscany Health Ecosystem)~\cite{THE} a collaborative framework that integrates healthcare providers, research institutions, and technology developers. The evaluation encompasses both clinical and technical assessments. The clinical validation will involve testing the predictive accuracy of AI/ML models, ensuring that risk stratification and scores to support decision-making provide clinically meaningful insights. The technical validation will focus on evaluating the platform and the telemonitoring kit in terms of data acquisition, transmission, usability, and integration with healthcare records. 
The outcomes from the empirical evaluation form the foundation for potential large-scale clinical trials, paving the way for broader adoption of the PrediHealth solutions.

\textbf{O5: Promote knowledge dissemination and adoption.} Disseminating the PrediHealth findings contributes meaningfully to both the scientific community and the broader healthcare ecosystem. To that end, we planned a dissemination strategy to maximize impact and foster the adoption of our solutions. We will actively engage in publishing research findings in high-impact scientific journals, presenting results at international conferences, and organizing targeted workshops with key stakeholders (\eg biomedical engineers). 


\renewcommand{\arraystretch}{1.2}  
\begin{table}[]
\caption{Description of the PrediHealth’s work packages}
\centering
\scalebox{0.8}{
\begin{tabular}{rp{0.3\textwidth}} 

\hline
\multicolumn{2}{c}{\textbf{WP1}: Coordination and Dissemination (UniSa DI)}
\\ \hline
\textbf{Objective:} & Coordination and dissemination of project results \\ 
\vspace{-8pt}
\textbf{Activity Breakdown:} & 
\vspace{-8pt}
 \begin{itemize}[left=0pt, labelsep=2pt]
    \item Definition and implementation of the project management model (M1-M15)
 \end{itemize}
\\ 
\vspace{-6pt}
\textbf{Deliverable(s):} & 
\vspace{-8pt}
\begin{itemize}[left=0pt, labelsep=2pt]
    \item Final report on coordination and dissemination (M15)
\end{itemize}
\\ \hline

\multicolumn{2}{c}{\textbf{WP2}: Study of Emerging Technologies and Solutions (UniSalento)} \\ \hline
\textbf{Objective:} & Study of emerging technologies with impact on the international healthcare landscape \\ 
\vspace{-8pt}
\textbf{Activity Breakdown:} & 
\vspace{-8pt}
 \begin{itemize}[left=0pt, labelsep=2pt]
    \item Scouting of innovative solutions for medical devices and environmental sensors (M1-M5)
     \item Analysis of Telemedicine technologies and solutions for the management of chronic patients (M1-M5)
 \end{itemize}
\\ 
\vspace{-6pt}
\textbf{Deliverable(s):} & 
\vspace{-8pt}
\begin{itemize}[left=0pt, labelsep=2pt]
    \item Report on innovative solutions and context analysis (M5)
\end{itemize}
\\ \hline

\multicolumn{2}{c}{\textbf{WP3}: Design and Prototyping of the Solution (UniSalento)} \\ \hline
\textbf{Objective:} & Design and prototyping of the web-based IoT platform \\ 
\vspace{-8pt}
\textbf{Activity Breakdown:} &
\vspace{-8pt}
 \begin{itemize}[left=0pt, labelsep=2pt]
    \item Definition of technical specifications, context analysis, involved actors, and organizational model (M4-M8)
     \item Design and prototyping of the overall web-based IoT platform solution (M6-M15)
 \end{itemize}
\\ 
\vspace{-6pt}
\textbf{Deliverable(s):} &
\vspace{-8pt}
\begin{itemize}[left=0pt, labelsep=2pt]
    \item Web-based IoT platform prototype and technical report (M15)
\end{itemize}
\\ \hline

\multicolumn{2}{c}{\textbf{WP4}: Definition and Training of the Predictive Model (UniSa DipMed)} \\ \hline
\textbf{Objective:} & Data collection and training of the predictive model\\ 
\vspace{-8pt}
\textbf{Activity Breakdown:} & 
\vspace{-8pt}
 \begin{itemize}[left=0pt, labelsep=2pt]
    \item Analysis of available AI indicators for decision support (M8-M11)
     \item  Development and prototyping of predictive models for patient trajectory (M9-M12)
     \item Pilot setup for data collection infrastructure (M11-M15)
 \end{itemize}
\\ 
\vspace{-6pt}
\textbf{Deliverable(s):} & 
\vspace{-8pt}
\begin{itemize}[left=0pt, labelsep=2pt]
    \item Prototyped predictive model and final training report (M15)
\end{itemize}
\\ \hline

\multicolumn{2}{l}{\scriptsize \textsuperscript{a} “M” stands for “Month”} 

\end{tabular}
}
\label{tab:WPs}
\end{table}

\section{Methodology and Structure} \label{sec:methodology}

Given the operational objectives above, the project is organized into four Work Packages (WPs).
Details on the WPs are reported in Table~\ref{tab:WPs}. Specifically, WP1, led by the Department of Computer Science at the University of Salerno (UniSA DI), addresses O5 and focuses on project coordination and dissemination of results throughout the entire project duration. WP2, coordinated by the University of Salento (UniSalento), covers the first five months and focuses on studying emerging technologies with potential healthcare applications at the international level. To some extent, this WP is cross-cutting to O1, O2, and O3 because each of these objectives required a specific examination of the state of the art and practice. WP3, also led by UniSalento, addresses O1 and O2. It begins in the fourth month and is dedicated to the deployment of a web-based IoT platform and a telemonitoring kit for HF patients. Finally, WP4, managed by the Department of Medicine at the University of Salerno (UniSA DipMed), addresses O3 and O4. It is centered around the data collection, the training of the predictive model, and the definition of scores for HF risk factors. The implemented research solutions are empirically evaluated in WP3 and WP4, and then both these WPs address O4. The industrial partner of PrediHealth contributed to the implementation aspects of WP3 and WP4, \ie the technical WPs of the project. A description of the project WPs follows.

\textbf{WP1: Coordination and dissemination of results.} It spans the full project duration and includes the definition and implementation of the project management model and the definition of strategies for the dissemination of the scientific results produced in the project. 

\textbf{WP2: Study of emerging technologies and solutions.} Its goal is to scout innovative and advanced medical devices and environmental sensors for the telemonitoring kit for HF patients. Additionally, it involves analyzing telemedicine platforms and web-based IoT platforms, technological solutions for integrated chronic patient management, the critical issues of current SC monitoring platforms, and AI/ML-powered predictive models for the early detection of HF diseases and risk stratification patients. 

\begin{table*}[t]
\vspace{-5mm}
    \renewcommand{\arraystretch}{2}
    \setlength{\tabcolsep}{6pt}
    \scriptsize
    \begin{center} 
    \caption{Comparison of IoT Platforms for PrediHealth}
    \label{tab:IoTPlatforms}
        \scalebox{0.7}{
        \begin{tabular}{cccccc}
            \hline  
            \textbf{Platform} & \textbf{IoT Protocols} & \textbf{Security} & \textbf{Data Visualization} & \textbf{External Integration} & \textbf{Applicability in Healthcare} \\ \hline
            \rule{0pt}{4ex}
            
            \textbf{ThingSpeak ~\cite{thingspeak}} & 
            MQTT, HTTP
            & API Authentication, SSL/TLS Encryption & Customizable Dashboards and Graphs & High & Medium, ~\cite{PHMSThingSpeak} \\ 
            
            \textbf{Fiware ~\cite{fiware}} &  MQTT, HTTP, CoAP, LWM2M, LoRaWAN, Sigfox, BLE & SSL/TLS Encryption, OAuth2, RBAC, 2FA & Customizable Dashboards, Graphs and Maps & High & High, ~\cite{fiwareEhealth, fiwareHealthBooklet, fiwareHospital}  \\ 
            
            \textbf{Mobius~\cite{mobius}} & MQTT, CoAP, HTTP & SSL/TLS, OAuth2 & Customizable Dashboards and Graphs & High & Medium \\ 
            
            \textbf{ThingsBoard ~\cite{thingsboard}} & MQTT, HTTP, CoAP, LWM2M, LoRaWAN, BLE & OAuth2, SSL/TLS Encryption, RBAC & Customizable Dashboards, Graphs and Maps & High & Medium, ~\cite{thingsboardHealthcare} \\ 
            
            \textbf{IoTivity ~\cite{iotivity}} & CoAP, MQTT & SSL/TLS Encryption, OAuth2, RBAC & Status Monitoring, Logs & Medium & Low \\
            
            \textbf{Californium ~\cite{californium}} & CoAP & DTLS Encryption, X.509 Authentication & Customizable Graphs, Status Monitoring & Medium & Low \\ 
            
            \textbf{Edge X Platform ~\cite{edgex}} & MQTT, HTTP, CoAP, LWM2M & TLS Encryption, OAuth2, ACL & Customizable Dashboards and Graphs & High & Medium \\ 
            
            \textbf{Eclipse Kapua ~\cite{kapua}} & MQTT, CoAP, HTTP, AMQP & SSL Encryption, RBAC & Customizable Dashboards and Graphs & High & Medium \\ 
            \hline
            
            
            
            
%
            
            
        \end{tabular} 
        }
    \end{center}
    \vspace{-5mm}
  
\end{table*}

\textbf{WP3: Design and prototyping of the solution.} This WP involves defining the technical specifications, performing context analysis, and setting the organizational model, along with the characterization of data collected from IoT devices and sensors. Additionally, the WP involves the design and prototypical implementation of the proposed solution infrastructure (\ie integrated web-based IoT platform and telemonitoring kit).
WP2 is closely linked to WP3, as the findings from the study on emerging technologies and solutions may inform the selection and customization of existing open-source platforms. The solutions implemented in WP3 will be validated through pilot projects or case reports with selected patients. 

\textbf{WP4: Definition and training of the predictive model.} This WP aims to study and define AI/ML-powered models to facilitate the stratification of patients at HF risk. Our primary goal was to facilitate the utilization of publicly available datasets (if any) to train AI/ML-powered models, ensuring robust and data-driven development within the project. As an alternative, we planned to use datasets of the THE ecosystem. Additionally, WP4 focuses on the prototypical development and assessment of predictive models and algorithms. WP4  also aims to define thresholds and scores for HF risk factors when using the telemonitoring kit.

\section{Result} \label{sec:results}
In this section, we report the most important preliminary results achieved in the early stage of PrediHealt arranged according to O1, O2, and O3 (\ie the objectives for which we obtained some preliminary results). 

\subsection{O1 - Interoperable Web-Based IoT Platform}

We studied and evaluated existing open-source web-based IoT platforms, considering scientific/gray literature (\eg technical papers, industry white papers, and regulatory documents) to ensure a thorough assessment of available technologies. We highlighted several potential candidates. 
The approach used for the selection of the web-based IoT platform and the focus on open-source solutions ensured community support and continuous updates as well as a solid and adaptable foundation while optimizing development time and costs. 


The selection process prioritized platforms that integrate with wearable devices and environmental sensors while ensuring compliance with evolving regulations and technological advancements. Platforms that lacked active maintenance or regular updates were excluded, as outdated solutions pose significant risks in terms of security, compatibility with modern devices, and adaptability to AI-driven predictive models. In contrast, modern and actively supported platforms with strong IoT integration were deemed more suitable for the project’s long-term sustainability. Further details on the selection process are omitted due to space constraints and as they fall outside the main goal of this paper (introducing PrediHealth and its ongoing results). In Table~\ref{tab:IoTPlatforms}, we report the final set of platforms from which we selected FIWARE~\cite{fiware}. 

FIWARE is a web-based IoT platform supported by the European Union~\cite{fiware,fiwareEU} and released under a permissive open-source license (BSD License). Backed by an active open-source community, FIWARE remains up-to-date. It is widely adopted in healthcare, with proven applications in hospital automation and real-time patient monitoring. In addition, FIWARE compliance with European digital health regulations provides strategic advantages~\cite{fiwareEhealth}, including potential funding opportunities. The modular architecture of FIWARE, based on Generic Enablers, allows customized remote health monitoring solutions. Scalability is key, as PrediHealth requires real-time processing of growing IoT data while maintaining high performance. FIWARE’s compatibility with Docker and Kubernetes supports high scalability and enhances deployment flexibility across cloud and edge environments. FIWARE, through the Context Broker, efficiently manages real-time data from IoT sensors and wearables. It also integrates advanced data processing tools for distributed computing, big data analytics, and ML, enabling AI-driven predictive healthcare. FIWARE is designed with a strong focus on security, incorporating multiple layers of protection to ensure data integrity, confidentiality, and controlled access. It implements robust authentication mechanisms through OAuth2, enabling secure authorization and identity verification for users and applications. Additionally, XACML-based access control policies allow for fine-grained permission management, ensuring that only authorized entities can access specific resources based on predefined rules. These features let FIWARE be compliant with community security standards and regulations. FIWARE also supports diverse IoT protocols, including MQTT, LoRaWAN, SigFox, OPC-UA, and Lightweight M2M, ensuring seamless device integration, interoperability, and compliance with healthcare standards. We designed and implemented an API to expose healthcare data in the HL7 FHIR format by leveraging the FIWARE's HL7 Smart Data Model~\cite{HL7SmartDataModel}.


\subsection{O2 - Telemonitoring Kit for HF Patients}

The design and the development of the telemonitoring kit for HF patients is based on the integration of environmental sensors and advanced (wearable and not) medical devices. As for the environmental sensors, monitoring environmental parameters is essential because parameters including air quality, temperature, and humidity complement the core vital signs in assessing cardiovascular health~(\eg \cite{khraishah2024understanding,singh2024heat}). In particular, air quality monitoring was included due to the established link between air pollution and cardiovascular risks. Khraishah \etal~\cite{khraishah2024understanding} showed that exposure to pollutants contributes to metabolic and CVDs and that wildfire-related air pollution exacerbates cardiovascular conditions. Similarly, temperature and humidity monitoring were prioritized given their impact on cardiovascular mortality, as extreme heat has been linked to increased cardiovascular-related deaths~\cite{singh2024heat}. We also considered environmental noise, given its association with hypertension and CVDs~\cite{basner2014auditory,altura1992noise,peterson1981noise,munzel2024transportation}. 
We conducted a study on various environmental sensors. The results indicated that, although there were differences in the characteristics of the sensors, none of them outperformed the others. Consequently, the selection process prioritized factors such as cost-effectiveness, measurement accuracy, and adherence to European manufacturing standards. This approach aligns with the requirements for projects funded by the National Recovery and Resilience Plan (PNRR), which suggests that the hardware used must be compliant with European standards/certifications. Table~\ref{tab:sensor_precision} shows the selected sensors and their technical specifications.

\begin{table}[]
\caption{Environmental Sensors}
\label{tab:sensor_precision}
\centering
\scalebox{0.8}{
\scriptsize
\begin{tabular}{p{3.3cm}p{5.8cm}}
\hline
\textbf{Sensor} & \textbf{Technical specifications} \\ \hline

\textbf{Sensirion SPS30} \cite{SensirionSPS30}\newline Air Quality and Control & 
\begin{minipage}[t]{\linewidth}
\begin{itemize}[left=0pt,nosep, after=\strut,label=-]
        \item Range: $0 - 1000 \, \mu\text{g/m}^3$ (PM1.0, PM2.5, PM4, PM10)
    \item Precision:
    \begin{itemize}
        \item PM1, PM2.5: $\pm [5 \, \mu\text{g/m}^3 + 5\% \, \text{m.v.}]$
        \item PM4, PM10: $\pm 25 \, \mu\text{g/m}^3$
    \end{itemize}
    \item Lifetime: $> 10$ years
    \item Size: $41 \times 41 \times 12 \, \text{mm}$
    \item Supply Voltage: $4.5 - 5.5 \, \text{V}$
    \item Current: $55 - 80 \, \text{mA}$
    \item Interfaces: I\textsuperscript{2}C, UART
\end{itemize}
\end{minipage}
\\
\hline

\textbf{Sensirion SCD41} \cite{SensirionSCD41}\newline CO$_2$ & 

\begin{minipage}[t]{\linewidth}
\begin{itemize}[left=0pt,nosep, after=\strut,label=-]
          \item CO$_2$ Output Range: $0 - 40,000$ ppm
    \item CO$_2$ Measurement Range: $400 - 5,000$ ppm
    \item CO$_2$ Accuracy: 
    \begin{itemize}
        \item $ \pm 50.0 \, \text{ppm} \pm 2.5\% \, \text{m.v.}$
        \item $ \pm 40.0 \, \text{ppm} \pm 5.0\% \, \text{m.v.}$
    \end{itemize}
    \item Response time\textbf{ ($\tau_{63\%}$)}: 60 s
    \item Supply Voltage: $2.4 - 5.5$ V
    \item Current: $15$ mA (avg), $205$ mA (max)
    \item Temperature Range: $-10 - 60$°C
    \item Size: $10.1 \times 10.1 \times 6.5$ mm
\end{itemize}
\end{minipage}
\\ \hline

\textbf{Sensirion SHT40i-BD1B} \cite{SensirionSHT40I}\newline Temperature & \begin{minipage}[t]{\linewidth}
\begin{itemize}[left=0pt,nosep, after=\strut,label=-]
         \item Temperature Accuracy: $\pm 0.2^\circ\text{C}$
    \item Temperature Range: $-40^\circ\text{C}$ to $+125^\circ\text{C}$
    \item Humidity Accuracy: $\pm 2\%$ RH
    \item Operating Humidity Range: $0$ to $100\%$ RH
    \item Response Time (Temperature): $2$ s
    \item Response Time (Humidity): $4$ s
    \item Supply Voltage: $2.3$ to $5.5$ V
    \item Average Supply Current: $21$ $\mu$A
    \item Interfaces: I\textsuperscript{2}C
    \item Size: $1.5 \times 1.5 \times 0.5$ mm
\end{itemize}
\end{minipage}
\\ \hline

\textbf{Sensirion STS31-DIS} \cite{SensirionSTS31}\newline Humidity & \begin{minipage}[t]{\linewidth}
\begin{itemize}[left=0pt,nosep, after=\strut,label=-]
         \item Temperature Accuracy: $\pm$0.2°C
    \item Operating Temperature Range: -40°C to +125°C
    \item Response Time (Temperature): 2 s
    \item Supply Voltage: 2.15 - 5.5 V
    \item Average Supply Current: 1.7 $\mu$A
    \item Max. Supply Current: 1500 $\mu$A
    \item Interfaces: I²C
    \item Size: 2.5 x 2.5 x 0.9 mm
    \item Humidity Accuracy: $\pm$2\% RH
    \item Operating Humidity Range: 0 - 100\% RH
    \item Humidity Response Time: 4 s
    \item Humidity Long-Term Drift: $<0.2$\% RH/year
\end{itemize}
\end{minipage}
\\ \hline

\textbf{Knowles SPK0641HT4H-1}
\cite{Knowles_SPK0641HT4H}
\newline Noise Pollution & \begin{minipage}[t]{\linewidth}
\begin{itemize}[left=0pt,nosep, after=\strut,label=-]
        \item Distortion: 2.2\% at 115 dB SPL
    \item SNR: 64.5 dB(A)
    \item Frequency Response: $20 \, \text{Hz}$ - $20 \, \text{kHz}$
    \item Voltage: $1.6 \, \text{V}$ to $3.6 \, \text{V}$
    \item Sensitivity: $-27$ to $-25 \, \text{dBFS}$ (94 dB SPL)
    \item THD: 0.1\% at 94 dB SPL
    \item AOP: $120 \, \text{dB SPL}$
    \item Low Power: $230 \, \mu\text{A}$
    \item PDM Output
    \item Omnidirectional
    \item Voltage Range: $-0.3 \, \text{V}$ to $+5.0 \, \text{V}$
    \item Operating Temperature: $-40^\circ \text{C}$ to $+100^\circ \text{C}$
\end{itemize}
\end{minipage}
\\ \hline

\textbf{Blue Line Gen 2 Bosch} \cite{BoschBlueLineGen2}\newline  Motion Detector & 
\begin{minipage}[t]{\linewidth}
\begin{itemize}[left=0pt,nosep, after=\strut,label=-]
     \item Power: 9V DC - 15V DC, 10mA.
    \item Temperature: G/GE: -20°C to +55°C, HE: +5°C to +40°C. 
    \item Humidity: 0-95
    \item Interference: Immunity up to 2 GHz
    \item Dimensions: 105 x 61 x 44 mm, white ABS plastic.
\end{itemize}
\end{minipage}
\\ \hline
\end{tabular}
}
\end{table}

Also, continuous weight monitoring is crucial for HF patients, as fluid retention can indicate worsening HF, as well as physical activity, as regular exercise has been shown to prevent and manage CVDs~\cite{schuler2013role,nieman1999exercise}. The kind of data gathered in PrediHealth with medical devices were chosen because they serve as critical indicators for detecting early signs of hemodynamic instability, autonomic dysfunction, or respiratory distress—key factors in managing HF
\cite{zeid2024heart,prieto2022wearable}.
More specifically, body temperature is crucial for identifying potential infections or inflammatory processes, which could signal a deterioration in the patient’s condition. Heart rate is a cardiovascular indicator that, in case of any abnormal increase or decrease, points to potential cardiac stress or arrhythmias. HRV, which reflects autonomic function, is particularly valuable for assessing the balance between the sympathetic and parasympathetic nervous systems, with low HRV being associated with an increased risk of cardiovascular events. SpO$_2$ is an important marker of respiratory function, with low levels indicating potential respiratory failure or pulmonary edema, which is common in HF patients. Blood pressure, both systolic and diastolic, is important for monitoring hypertension, a significant risk factor for HF. Fluctuations in blood pressure may indicate a worsening cardiac function or the development of pulmonary hypertension. Additionally, ECG data provide detailed insights into the heart’s electrical activity, allowing for the detection of arrhythmias, ischemia, and other abnormalities that may require immediate intervention. The integration of these kinds of data with physiological information supports the early detection of potential health deterioration and allows for proactive interventions.


We focused on devices that ensure high measurement accuracy, seamless integration with our IoT platform, and compliance with medical standards (\eg HL7 FHIR). A comprehensive analysis of commercially available devices was conducted, encompassing smartwatches, smartbands, sensor-integrated smart garments, and weight-monitoring devices, all of which could play a role in HF telemonitoring. We considered scientific literature and gray literature to ensure a thorough assessment of available technologies. The selected devices were assessed based on multiple criteria, including their ability to provide real-time physiological and behavioral data, interoperability with third-party healthcare applications, and accessibility through APIs or SDKs. Due to space constraints, Table~\ref{tab:devices} shows only the selected devices. Specifically, we selected the Withings ScanWatch 2 as the primary wearable device. This choice was driven by its ability to record a single-lead ECG directly from the wrist, enabling early detection of arrhythmias such as atrial fibrillation. This device also provides continuous monitoring, \eg of heart rate and SpO$_2$, offering a comprehensive cardiovascular assessment. A key advantage of ScanWatch 2 is its integration with IoT platforms via Withings' well-documented APIs. These APIs ensure secure access to real-time health data (\eg ECG traces), facilitating efficient data collection and analysis. Moreover, ScanWatch 2 meets stringent medical certification standards, ensuring measurement accuracy and compliance with regulatory requirements. Compared to other smartwatches, ScanWatch 2 offers a balance between advanced medical-grade features and ease of integration, making it an optimal solution for telemonitoring. For similar reasons, we chose the Withings Body+ scale. Further details on the selection process of the scale are not provided for space reasons.

\begin{table}[t]
\caption{Medical Devices}
\centering
\scalebox{0.68}{
\begin{tabular}{p{3cm}p{3cm}p{2.5cm}p{2.5cm}}
\hline 
 \textbf{Device} &  \textbf{Available \newline Measurements} &  \textbf{Data Access \newline (API/SDK)} & \textbf{Battery Life \newline(days)}
\\ \hline
\centering \textbf{Withings ScanWatch 2} & 
\begin{minipage}[t]{\linewidth}
\begin{itemize}[left=0pt,nosep,after=\strut,label=-]
    \item ECG 
    \item Heart Rate
    \item Body Temperature
    \item SpO2
    \item Physical Activity
    \item Sleep Monitoring
\end{itemize}
\end{minipage} &
Yes &
30
\\ \hline

\centering \textbf{Withings Body+} & 
\begin{minipage}[t]{\linewidth}
\begin{itemize}[left=0pt,nosep, after=\strut,label=-]
  \item Weight
  \item Body Mass Index
  \item Fat mass
  \item Muscle mass
  \item Bone mass
  \item Water percentage
\end{itemize}
\end{minipage}&
Yes &
540
\\ \hline

\end{tabular}} \label{tab:devices}
\end{table}

\begin{table}[t]
    \caption{Features used in the model}
    \label{tab:features}
    \centering
    \scriptsize
    \renewcommand{\arraystretch}{1} 
    \setlength{\tabcolsep}{1pt}       
    \scalebox{0.80}{
    \begin{tabularx}{\columnwidth}{p{1.7cm}p{7cm}}
        \hline
        \multicolumn{2}{c}{\textbf{Clinical and Echocardiographic Features}} \\ 
        \hline
        \textbf{Feature} & \textbf{Description} \\ 
        \hline
        \multicolumn{2}{c}{\textbf{Clinical Features}} \\ 
        \hline
        Diagnosi & Primary diagnosis (Divided into Primary and Secondary if two diagnoses are present) \\
        HFpEF & Heart Failure with Preserved Ejection Fraction \\
        EF & Ejection Fraction (\%) – Left ventricular pumping function \\
        NYHA & New York Heart Association (NYHA) class (I-IV), measures heart failure severity \\
        Age & Patient's age \\
        BMI & Body Mass Index \\
        Sex & Patient's sex (M/F) \\
        Hypertension & Hypertension (Yes/No) \\
        Dyslipidemia & Dyslipidemia (Yes/No) \\
        Diabetes & Diabetes mellitus (Yes/No) \\
        COPD & Chronic obstructive pulmonary disease (COPD) (Yes/No) \\
        BetaBlocc & Beta-blocker therapy (Yes/No) \\
        ACE\_SART & ACE inhibitors or Sartans therapy (Yes/No) \\
        AntiAldosterone & Aldosterone antagonists (e.g., spironolactone) (Yes/No) \\
        \hline
        \multicolumn{2}{c}{\textbf{Echocardiographic Features}} \\ 
        \hline
        PARETE POST & Left ventricular posterior wall thickness (mm) \\
        SETTO & Interventricular septum thickness (mm) \\
        LVES\_DIAM & Left Ventricular End-Systolic Diameter (mm) \\
        LVED\_DIAM & Left Ventricular End-Diastolic Diameter (mm) \\
        VDx (PARAST) & Right ventricular diameter (parasternal view) (mm) \\
        LVMI & Left Ventricular Mass Index (g/m²) \\
        ASx & Left atrial diameter or volume \\
        TAPSE & Tricuspid Annular Plane Systolic Excursion (mm), right ventricular function \\
        RS & Radial Strain, myocardial deformation measurement \\
        BBSx & Left Bundle Branch Block (Yes/No) \\
        BBDx & Right Bundle Branch Block (Yes/No) \\
        EF & Ejection Fraction (\%) – Repeated in this section \\
        NT-proBNP & N-terminal pro B-type Natriuretic Peptide (pg/mL), heart failure biomarker \\
        Creatinine & Blood creatinine level (mg/dL), renal function marker \\
        Glucose & Blood glucose level (mg/dL) \\
        FA & Atrial Fibrillation (Yes/No) \\
        Flutter & Atrial Flutter (Yes/No) \\
        PM & Pacemaker presence (Yes/No) \\
        Hb & Hemoglobin level (g/dL), anemia or oxygen transport marker \\
        \hline
    \end{tabularx}
    }
\end{table}

\subsection{O3 - Developing tools to support decision-making}
We implemented the predictive model for stratification using clinical and echocardiographic features. The used features are reported in Table~\ref{tab:features}. Since we did not find a public dataset to train such a model, we had to use a dataset of the THE ecosystem. This choice could have affected the external validity of the results because the performances of our model might differ in other populations with varying demographic profiles. The original dataset contained 1040 elements that became 357 after pre-processing. 
Further details on this phase are available in~\cite{cassieri2025predictivemodelschronicheart}.

The model architecture is based on an ensemble learning technique, specifically a variation of \textit{stacking}~\cite{Pavlyshenko_2018}. In standard stacking, multiple predictive models (often different from each other) are trained on the same set of features, and a meta-model then combines their outputs to generate the final prediction. Differently, 
we employed two specialized models trained separately: one using only clinical features and the other using only echocardiographic features. The outputs of these two models serve as inputs for the meta-model (a custom version of a voting classifier), which provides the final stratification of patients. 

The performance of our model was assessed using metrics that are commonly employed in the ML and medical domains. Specifically, we considered: (i)~Accuracy ($\frac{TP + TN}{TP + TN + FP + FN}$), (ii)~Precision ($\frac{TP}{TP + FP}$), (iii)~Sensitivity/Recall ($\frac{TP}{TP + FN}$), (iv)~F1-Score ($2 \times \frac{\text{Precision} \times \text{Sensitivity}}{\text{Precision} + \text{Sensitivity}}$), and (v)~Diagnostic Odds Ratio ($\frac{TP \times TN}{FP \times FN}$).
For the sake of clarity, TPs (True Positives) refer to patients correctly classified as people at HF risk, while TNs (True Negatives) refer to patients correctly identified as not at HF risk. Instead, FPs (False Positives) represent patients mistakenly classified as at risk, while FNs (False Negatives) represent patients mistakenly identified as not at risk.

\begin{table}[]
\centering
\caption{Medical evaluation metrics of the proposed model}
\label{tab:metrics}
\scalebox{0.8}{
\begin{tabular}{lc}
\hline
\textbf{Metric} & \textbf{Value}  \\
\midrule
Accuracy     & 78\%   \\
Precision    & 70\%   \\
Sensitivity  & 91\%   \\
F1-Score      & 79\%  \\
Diagnostic Odds Ratio & 20  \\
\hline
\end{tabular}
}
\end{table}

The experimental results of our model are summarized in Table~\ref{tab:metrics}. These results suggest that the model performs pretty well. Although an accuracy of 78\% could be considered moderate in some contexts, it is acceptable in the PrediHealth project given the priority of capturing patient at HF risk. The high sensitivity (91\%) ensures that nearly all HF risk patients are detected, while a relatively low precision (70\%) could represent an acceptable trade-off in our project. Although participation in the telemonitoring program could be expensive, it is better to include patients mistakenly classified as at risk (FPs) rather than not including patients mistakenly identified as not at risk (FNs) in such a program. The good balance between sensitivity and precision is shown by the obtained F1-Score value (79\%). The high Diagnostic Odds Ratio (20) confirms the robustness of the model to discriminate patients in at HF risk or not. Further information on the model is not provided for space constraints and because it is out of the scope of this paper and the initial results of its experimental assessment are available in~\cite{cassieri2025predictivemodelschronicheart}.

As for the thresholds/scores for risk factors for HF exacerbations when using telemonitoring kits, we chose them on clinical evidence and guidelines~\cite{Ehab368,CIR.0000000000001168,CIR.0000000000001062}. Specifically, threshold values considered at risk include SpO$_2$ $<$ 92\%, resting heart rate $>$ 100 bpm (tachycardia) or $<$ 50 bpm (bradycardia), and persistently low HRV, typically Standard Deviation of the Normal-to-Normal $<$ 20 ms. Early signs of decompensation can be inferred when body weight increases by more than 2 kg in three days, suggesting fluid overload. Other critical indicators include systolic blood pressure $>$ 140 mmHg or $<$ 90 mmHg, and diastolic pressure $>$ 90 mmHg or $<$ 60 mmHg, which may indicate hemodynamic imbalance. Finally, a respiratory rate $>$ 20 breaths/min and body temperature $>$ 37.5$^{\circ}$ or $<$ 36$^{\circ}$ may signal infection, hypoperfusion, or cardiovascular stress. Finally, a single-lead ECG, validated for atrial fibrillation detection by identifying irregular RR intervals and absent P waves, enables early recognition of AF, which is critical as it can impair cardiac output and increase the risk of decompensation and thromboembolic events. When risk factors occur, the IoT platform will notify both the patient and the appropriate healthcare infrastructure. This phase of PrediHealth is still ongoing.






\section{Final Remarks and Expected Impact}  \label{sec:conclusion}

The PrediHealth project is centered around the development of a web-based IoT application designed for remote Heart Failure (HF) patient monitoring. This innovative solution enables the continuous monitoring of vital signs and environmental parameters, thereby improving patient engagement and allowing for real-time oversight. By leveraging AI/ML-powered algorithms, we provide support to clinical decision-making (\eg patient stratification), monitor disease, and proactively prevent exacerbations. 
With its holistic and data-driven approach to integrating clinical and environmental factors, PrediHealth has the potential to transform HF management, foster personalized care, and contribute to long-term healthcare sustainability. 

\section*{Acknowledgment}
This project has been financially supported by the European Union NEXTGenerationEU project and by the Italian Ministry of the University and Research MUR (M4C2) as part of the project ``Tuscany Health Ecosystem'' - THE - Spoke 10 - CUP - J13C22000420001. One of the authors involved in the presented research is funded by the project ``DHEAL – COM- Digital Health Solutions in Community Medicine" under the Innovative Health Ecosystem (PNC) - National Recovery and Resilience Plan (NRRP) program funded by the Italian Ministry of Health.

\bibliographystyle{IEEEtran}
\bibliography{bibliography}
\vspace{12pt}

\end{document}